\documentclass[aps,prb,showpacs,twocolumn,letterpaper]{revtex4}
\usepackage{graphicx}
\usepackage{amsmath}
\begin{document}

\title{Hydrogen Dynamics in Superprotonic CsHSO$_{4}$}

\author{Brandon C. Wood}
\author{Nicola Marzari}
\affiliation{Department of Materials Science and Engineering, Massachusetts 
Institute of Technology, Cambridge, MA 02139}

\begin{abstract}

We present a detailed study of proton dynamics in the hydrogen-bonded 
superprotonic conductor CsHSO$_4$ from first-principles molecular dynamics
simulations, isolating the subtle interplay between the dynamics of the O--H 
chemical bonds, the O$\cdots$H hydrogen bonds, and the SO$_4$ tetrahedra in
promoting proton diffusion. We find that the Grotthus mechanism of proton 
transport is primarily responsible for the dynamics of the chemical bonds, 
whereas the reorganization of the hydrogen-bond network is dominated by rapid 
angular hops in concert with small reorientations of the SO$_4$ tetrahedra. 
Frequent proton jumping across the O--H$\cdots$O complex is countered by a 
high rate of jump reversal, which we show is connected to the dynamics of the 
SO$_4$ tetrahedra, resulting in a diminished CsHSO$_4$/CsDSO$_4$ isotope 
effect. We also find evidence of multiple timescales for SO$_4$ reorientation 
events, leading to distinct diffusion mechanisms along the different crystal 
lattice directions. Finally, we employ graph-theoretic techniques to 
characterize the topology of the hydrogen-bond network and demonstrate a 
clear relationship between certain connectivity configurations and the 
likelihood for diffusive jump events.

\end{abstract}


\maketitle

\section{Introduction and motivation}

Despite extensive efforts to realize the hydrogen economy, key technological 
hurdles to the widespread adoption of fuel-cell technology remain. One of 
the several challenges is the discovery and optimization of proton-conducting 
electrolyte materials for mid- to high-temperature use. Recently, a great deal 
of research effort has focused on anhydrous solid-state materials, in part 
because these offer greater flexibility in their range of operating 
temperatures than materials containing liquid water. Among the promising 
candidates are various derivatives of CsHSO$_{4}$, which has been shown to 
exhibit high ionic conductivity ($>10^{-2}$~($\Omega\cdot$cm)$^{-1}$)
\cite{baranov82} at target operating temperatures. Fuel-cell operation using 
electrolytes based upon CsHSO$_{4}$ and similar materials has already been 
successfully demonstrated in the laboratory.\cite{haile01,haile04} A complete 
theoretical picture of the detailed atomistic mechanisms involved in proton 
transport in these materials is highly desirable, since such knowledge 
would be useful in directing future research efforts in the optimization and 
adaptation of new materials. Previous molecular dynamics studies\cite{munch95,
munch96,goddard05} of CsHSO$_{4}$ based on fitted interatomic potentials
have aided in highlighting the basic phenomenology of proton transport, but 
such investigations are unable to capture the full complexity of hydrogen 
bonding and electronic interactions, particularly in a dynamic environment 
that features rapid bond breaking and forming. The work of Ke and 
Tanaka\cite{ke04a,ke04b} incorporated first-principles methodologies, but 
their analysis was grounded in static rather than dynamics calculations. The 
present study aims to elucidate the detailed atomistic pathways and mechanisms 
involved in hydrogen diffusion in superprotonic CsHSO$_{4}$ using 
first-principles molecular dynamics.

\begin{figure}
	\centering
	\includegraphics[width=2.3in]{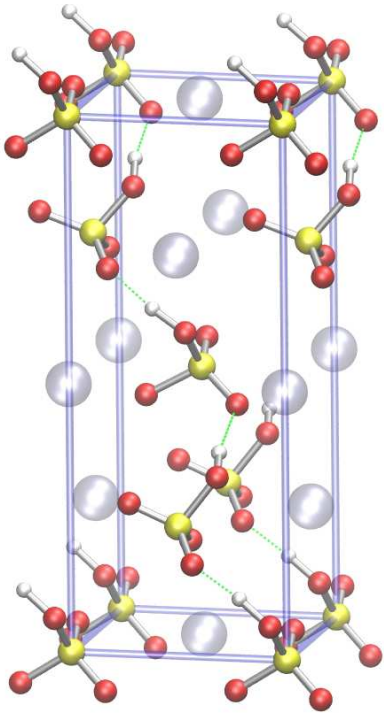}
	\caption{(Color online) Structure of the conventional unit cell of 
	Phase-I CsHSO$_4$. Hydrogen atoms are shown in white, oxygen in red, 
	sulfur in yellow, and cesium in blue. Hydrogen bonds are denoted by 
	broken green lines.}
	\label{fig:cshso4_structure}
\end{figure}

CsHSO$_{4}$ was the first known crystalline material to exhibit both hydrogen 
bonding and superprotonic behavior and has among the higher ionic 
conductivities of the known solid-acid materials. The room-temperature phase 
is monoclinic and features a static, well-defined network of hydrogen bonds. 
The higher-temperature superprotonic phase (usually designated Phase I) 
possesses a body-centered tetragonal structure and is stable above 
414K.\cite{baranov82} The unit cell of this phase is depicted in 
Fig.~\ref{fig:cshso4_structure} and consists of a lattice of SO$_4$ 
tetrahedra, each bonded to a hydrogen via an O--H chemical bond. Each 
chemically bonded hydrogen also forms an O$\cdots$H hydrogen bond with an 
oxygen of a neighboring SO$_4$ tetrahedron. The resulting hydrogen-bond 
network becomes dynamic above the transition temperature and can visit a 
number of distinct topologies, owing to four possible oxygen binding sites for 
each SO$_4$ node in the network. The reigning view in the literature
\cite{phamthi85,merinov87,kreuer97,belushkin98,zetterstrom99} is that 
long-range proton transport in superprotonic CsHSO$_{4}$ occurs as the net 
result of two separate mechanisms: first, the reorientation of the 
hydrogen-bond network by rapid rotations of the sulfate tetrahedra; and 
second, the hopping of the proton between oxygens of neighboring tetrahedra 
across the O--H$\cdots$O complex. The second step has generally been 
considered rate limiting and is thought to occur at frequencies of the order 
10$^{-9}$~s$^{-1}$, whereas the first is expected to happen more frequently 
by at least two orders of magnitude.\cite{badot89,belushkin92} This two-step 
process is often referred to collectively as the \emph{Grotthus mechanism}.

\section{Method}

We performed Car-Parrinello molecular dynamics simulations\cite{cp85} of
Phase-I CsHSO$_{4}$ in the canonical \textit{NVT} ensemble at temperatures of 
550~K, 620~K, and 750~K (superheated), with temperatures maintained by means of
a Nos\'{e}-Hoover thermostat\cite{nose84a,nose84b,hoover85}. Each simulation 
covered 25~ps of thermalized dynamics following 5~ps of equilibration. This 
length of time proved sufficient for sampling several hundred to a thousand 
jump events, making stastical inferences possible. In each case, our supercell 
was comprised of 112 atoms (sixteen complete CsHSO$_{4}$ units). Simulations 
were performed in a plane-wave basis using standard-valence ultrasoft 
pseudopotentials for hydrogen, oxygen, and sulfur; a 
$6s^{0.5}5d^{0.05}6p^{0.05}$ norm-conserving pseudopotential with nonlinear 
core correction for cesium; and the Perdew-Burke-Ernzerhof 
exchange-correlation functional\cite{pbe96}. All pseudopotentials are
obtainable from the Quantum-ESPRESSO website\cite{espresso}. Cutoffs of 25~Ry 
and 150~Ry were used for the wavefunctions and charge density, respectively. 
The fictitious Car-Parrinello mass was $\mu=700$ with $\Delta t=7.5$~au, and 
the lattice parameter was chosen based on the experimental value just above 
the superionic transition temperature, taken from 
Ref.~\onlinecite{belushkin91a}.

\begin{figure}
	\centering
	\includegraphics[width=3in]{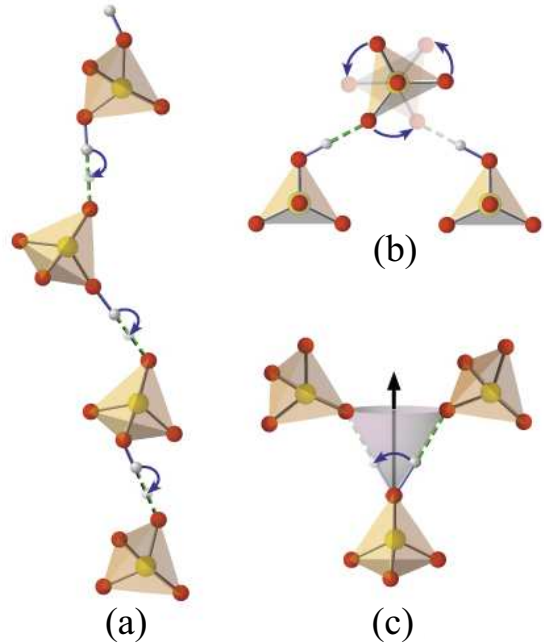}
	\caption{(Color online) Schematic depiction of (a) a sequence of 
	chemical-bond jumps nucleated by the formation of an H$_2$SO$_4$ 
	defect in the uppermost tetrahedron; (b) a hydrogen-bond network 
	change induced by rotation of a host SO$_4$ tetrahedron; and (c) a 
	hydrogen-bond network change resulting from a direct hydrogen-bond hop 
	with little or no rotation of the host SO$_4$ tetrahedron. The color 
	scheme follows that of Fig.~\ref{fig:cshso4_structure}, with final 
	configurations in jumping events shown as semi-transparent.}
	\label{fig:hopping}
\end{figure}

In presenting the results of our simulations, we have divided the dynamics into
categories of chemical-bond dynamics and hydrogen-bond dynamics. We include in 
our definition of chemical-bond dynamics any breaking or forming of O--H 
chemical bonds by Grotthus-type hopping of a proton between oxygens of 
neighboring tetrahedra. Any change in the hydrogen-bond network structure 
resulting from breaking or forming O$\cdots$H hydrogen bonds that does 
\emph{not} also involve breaking or forming O--H chemical bonds is considered 
in the category of hydrogen-bond dynamics. For additional clarity, 
Fig.~\ref{fig:hopping} offers schematic representations of sample jump events
from each category of dynamics.

\section{Chemical-bond dynamics}

\begin{figure}
	\centering
	\includegraphics[angle=270]{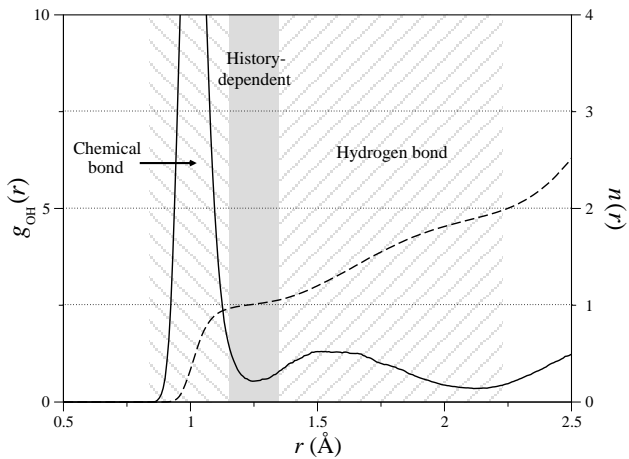}
	\caption{The oxygen-hydrogen radial pair distribution function
	$g_{\mathrm{OH}}(r)$ (solid line, left axis) and the coordination 
	number $n(r)$ (dashed line, right axis), calculated from a simulation 
	at 620~K. Regions classified as chemical and hydrogen bonds are 
	delineated, as well as the intermediate region for which the 
	history-dependent definition was employed.}
	\label{fig:g_oh}
\end{figure}

An O--H chemical bond was defined by considering interactions with oxygens 
within a cutoff distance of $R_{\mathrm{OH}}<1.15$~\AA, which represents the
initial separation between the first and second coordination peaks of the 
calculated oxygen-hydrogen radial pair distribution function (RDF), displayed
in Fig.~\ref{fig:g_oh}. Classification as chemical or hydrogen bond proved 
more difficult for O--H pairs separated by an intermediate distance 
$1.15\leq R_{\mathrm{OH}}<1.35$~\AA\ due to an inherent difficulty in 
resolving the overlap in the first two RDF peaks in that range. The ambiguity
is also noticeable upon examination of the coordination number $n(r)$, which is
nearly flat in this region. For such O--H pairs, we instead employed a 
history-dependent definition, basing the bond category on the last visited
unambiguous bonding regime. Under this definition, an existent O--H chemical 
bond was considered broken only when $R_{\mathrm{OH}}\geq 1.35$~\AA; 
analogously, an existent O$\cdots$H hydrogen bond was considered broken only 
when $R_{\mathrm{OH}}<1.15$~\AA.

Within the Grotthus mechanism, local proton transfer via a series of
correlated jumps prompts changes in the chemical-bond structure. Such 
jumps are first nucleated by the formation of a metastable H$_{2}$SO$_{4}$ 
defect, which subsequently propagates along the network backbone, acting 
as a successive proton donor for neighboring tetrahedra at each stage. The
individual jumps are themselves short hops across a double-well potential 
barrier, where the two stable minima represent the O$\cdots$H and O--H 
distances and are separated by about 0.5~\AA. A single Grotthus hop therefore 
has the effect of swapping a chemical and a hydrogen bond, an action that is 
repeated as the proton transfer propagates across the hydrogen-bond network 
chain. This model of local proton transfer in superprotonic CsHSO$_4$, 
represented schematically in Fig.~\ref{fig:hopping}(a), is easily observable 
in our simulations. Indeed, just over half (51\%) of the chemical-bond jumps 
that we register at 620~K occur as a direct result of H$_{2}$SO$_{4}$ defect 
formation by the donation of a second proton from a neighboring tetrahedron, in
accordance with the Grotthus model. The remaining jumps are nucleated as a 
result of random local fluctuations in the bond structure.

\begin{figure}
	\centering
	\includegraphics[angle=270]{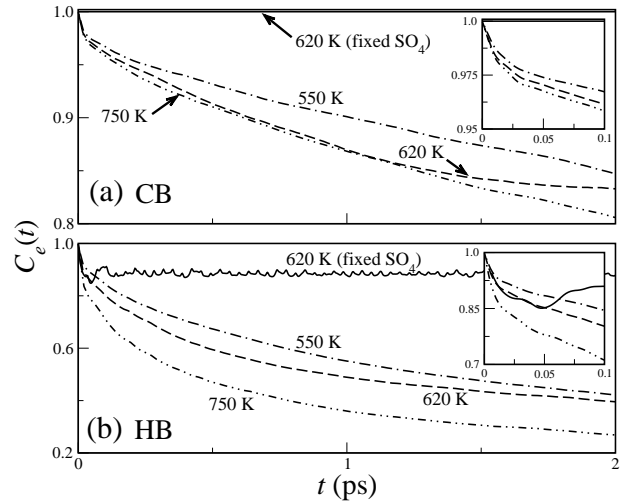}
	\caption{Autocorrelation functions for existence of (a) chemical and
	(b) hydrogen bonds after a time $t$ for simulations at 550~K, 620~K, 
	and 750~K, as well as at 620~K with fixed sulfate tetrahedra.}
	\label{fig:cbhb_autocorr}
\end{figure}

It is a straightforward process to track bond formation and annihilation, and 
we can define a time autocorrelation function for bond existence as 
$C_{e}(t)=\langle \alpha(0)\alpha(t) \rangle$, where $\alpha(t)$ is 1 if a 
particular type of bond exists between a given hydrogen-oxygen pair at time 
$t$, and 0 otherwise. In practice, we can improve our statistical sampling by 
averaging $C_{e}(t)$ over all available time intervals of length $t$ in the 
simulation. Using this quantity, we can obtain a detailed picture of the 
timescales of the chemical- and hydrogen-bond dynamics. The hydrogen-bond and 
chemical-bond existence autocorrelation functions are displayed in 
Fig.~\ref{fig:cbhb_autocorr}.  

Beyond about 20~fs, we observe a slow exponential decay in the chemical-bond
existence autocorrelation $C_e(t)$, with characteristic $1/e$ decay times in 
the 11--15~ps range. These values are recorded in Table~\ref{tab:cb_data}, 
along with average overall statistical frequencies of chemical-bond jumps for 
each temperature and the fraction of such jumps that subsequently reverse 
themselves. Here a chemical-bond jump is defined as a complete exchange of a 
chemical and a hydrogen bond across an O--H$\cdots$O complex. The surprising 
commonality of chemical-bond jump events is reconcilable with the slow decay 
rate only when one considers the extremely high rate of jump reversal, which 
averages around 85\% and features no significant variability with temperature.
The combination of frequent jumping and a high, temperature-independent 
reversal rate is suggestive of a potential energy surface featuring an 
especially shallow activation barrier.

At very short times ($<$10~fs), we observe a fast fall-off before 
transitioning to the slower decay regime. In this region, we are below the
timescale of any jump reversal subsequent to chemical-bond breaking and
forming, resulting in a much more rapid decay. The absence of any noticeable
high-frequency periodicity in Fig.~\ref{fig:cbhb_autocorr}(a) indicates that
jump reversal carries no preferred timescale. Rather, it is likely that the 
reversal probability is a consequence of stabilization or destabilization 
of the local potential energy surface from SO$_4$ tetrahedral reorientations.

\begin{table}
	\caption{Various quantities derived from a statistical analysis of the 
	chemical-bond dynamics at 550~K, 620~K, and 750~K. Featured columns 
	indicate (1) the characteristic decay time $\tau$ in an exponential 
	fit $Ae^{-(t/\tau)}$ of the long-time data in 
	Fig.~\ref{fig:cbhb_autocorr}(a); (2) the average overall observed 
	frequency $\nu_{\mathrm{c}}$ of chemical-bond jump events per ion; and 
	(3) the fraction of these events that subsequently reverse themselves.}
	\label{tab:cb_data}
\begin{tabular}
        {c|ccc}
	\hline\hline
	~\emph{Temperature} (K)~ & ~$\tau$ (ps)~ 
	& ~$\nu_{\mathrm{c}}$ (THz)~ & ~\emph{\% rev.}~ \\
	\hline
	550 & 15.2 & 0.58 & 83 \\
	620 & 12.5 & 0.80 & 85 \\
	750 & 11.2 & 0.94 & 85 \\
	\hline\hline
\end{tabular}
\end{table}

As an indicator of the potential effect of the motion of the SO$_{4}$ 
tetrahedra on the chemical- and hydrogen-bond dynamics, we have also chosen to 
run a second simulation at 620~K in which all ions except for the hydrogens 
were immobilized (denoted ``fixed-SO$_4$'' in Fig.~\ref{fig:cbhb_autocorr}). 
The configuration was chosen from a well-equilibrated timestep of the fully 
mobile simulation. Interestingly, all chemical-bond dynamics ceased in this 
simulation. This fact is particularly notable in light of previous 
investigations\cite{munch95,munch96,kreuer97} which postulated that the most 
important factor in inducing chemical-bond jumping is a reduction in the 
O--O distance across the O--H$\cdots$O complex due to SO$_4$ reorientation.
However, in our fixed-SO$_4$ simulation, not a single chemical-bond jump event 
was registered, despite the continuous presence of O--O distances as short as 
2.30~\AA\ across O--H$\cdots$O complexes. Notably, this value is approximately 
0.15~\AA\ shorter than the average O--O distance across O--H$\cdots$O 
complexes of hydrogens actively involved in chemical-bond jumps in the fully
mobile simulation. Moreover, in our fully mobile simulations, we find only a 
0.01~\AA\ difference between the average O--O distances across O--H$\cdots$O 
complexes of hydrogens involved in chemical-bond jumping events and ordinary
hydrogens not involved in jump events of any sort. We therefore conclude that 
the primary contribution of the SO$_4$ tetrahedra to chemical-bond jumping 
results from their vibrational or rotational dynamics rather than simply their 
instantaneous orientation. Also, despite the low barrier for Grotthus-style 
chemical-bond hopping, dynamic degrees of freedom connected solely to the 
hydrogens are nevertheless insufficient to permit chemical-bond breaking or 
forming, detailing the necessary role of the oxygen modes in that process. 

Our results reveal that the general chemical-bond jump frequency is relatively
high; moreover, it is of the same order as the hydrogen-bond dynamics (compare
Tables~\ref{tab:cb_data} and \ref{tab:hb_data}). This contrasts with the view 
of chemical-bond jumping as substantially rate limiting and differing in
timescale from the hydrogen-bond dynamics by two or more orders of magnitude. 
We instead find that the limiting factor in the \emph{effective} rate of 
chemical-bond jumps is the extraordinarily high rate of jump reversal, which 
we suggest is linked to the dynamics of the SO$_4$ tetrahedra. Yet even when
jump reversals are considered, our effective chemical-bond dynamics are 
significantly faster than the proposed nanosecond scale. Instead, our results 
are consistent with recent experiments\cite{mizuno04,hayashi04} pointing
to much faster chemical-bond dynamics than have hitherto been assumed, 
perhaps even on the picosecond scale.

It should be noted that in our analysis of jump reversals, we have considered 
only \emph{single} jumps that are subsequently reversed. In so doing, we have 
neglected the reversal of collective sequences of jump events. Such 
higher-order jump sequences are especially difficult to properly consider 
given the inherent time- and lengthscale limitations imposed by the 
first-principles methodology. It is expected that accounting for the potential 
reversal of such longer sequences would limit the number of counted 
``successful'' jumps (those contributing to macroscopic proton diffusion). The 
magnitude of any such limitation is difficult to predict, however.

\section{Hydrogen-bond dynamics}

For the purposes of this work, we have defined a hydrogen bond in terms of the 
oxygen-hydrogen distance alone, with the additional restriction that hydrogen 
bonding cannot involve oxygens attached to the same host sulfate tetrahedron 
as the hydrogen. Under this definition, the usual practice of restricting 
$\angle$O--H$\cdots$O had no appreciable effect on the counted hydrogen bonds
and was omitted for sake of simplicity. A hydrogen-bond maximum cutoff distance 
of $R_{\mathrm{OH}}<2.23$~\AA\ was chosen based on the distance for which the
coordination number $n(r)=2$, indicating the tail end of the second coordation 
peak (associated with H$\cdots$O) in the calculated oxygen-hydrogen RDF 
(Fig.~\ref{fig:g_oh}). The minimum cutoff of $R_{\mathrm{OH}}\geq1.35$~\AA\ 
was chosen based on the clear point of separation for the second RDF peak and
the end of the plateau region in the coordination number. As for the chemical 
bonds, we implement a history-dependent definition for categorizing bonds in 
the intermediate range of $1.15\leq R_{\mathrm{OH}}<1.35$~\AA.

Fig.~\ref{fig:cbhb_autocorr}(b) shows the bond-existence autocorrelation
function $C_e(t)$ for the hydrogen bonds in simulations at 550~K, 620~K, 750~K,
and for the ``fixed-SO$_4$'' simulation at 620~K in which all ions except the 
hydrogens are immobilized. At longer times, we observe an exponential decay of
the hydrogen bonds for the fully mobile simulations that far outpaces that of 
the chemical bonds. The graph also reveals that at short times ($<50$~fs; see 
figure inset), the hydrogen bond network in the fixed-SO$_4$ simulation 
remains very dynamic, approximately following the equivalent curve for the 
fully mobile system. However, $C_e(t)$ soon begins to oscillate around a fixed 
running average, indicating repeated visitation of a few alternating 
configurations. Interestingly, the overall frequency of hydrogen-bond breaking 
is actually greater for the fixed-SO$_4$ simulation than for the fully mobile 
simulation.

Table~\ref{tab:hb_data} contains these hydrogen-bond breaking frequencies, as 
well as likelihoods for reversal of hydrogen-bond network reorganization 
events. In addition to providing overall values, we have divided the 
hydrogen-bond dynamics into two categories based on the location of the newly
formed hydrogen bond with respect to its predecessor. Our first category
consists of hydrogen bonds transferred between oxygens of the same destination 
SO$_4$ tetrahedron; the second contains hydrogen bonds transferred between 
oxygens of neighboring SO$_4$ tetrahedra. In practice, higher temperatures 
generally show a greater preference for exchanges between oxygens of different 
tetrahedra than do lower temperatures, but in all cases, such exchanges 
outnumber those between oxygens of the same tetrahedron by a margin of two or 
three to one. Fixing the SO$_4$ tetrahedra pushes that margin even further.

\begin{table*}
	\caption{Various quantities derived from a statistical analysis of the 
	hydrogen-bond dynamics at 550~K, 620~K, 750~K, and for the fixed-SO$_4$
	simulation at 620~K. The data is divided into statistics for 
	hydrogen-bond exchanges between oxygens of the same SO$_4$ 
	tetrahedron, those of different SO$_4$ tetrahedra, and overall totals 
	for either type of exchange. Featured columns include (1) the fraction 
	of total hydrogen-bond exchanges representing a particular class of 
	exchange; (2) the average overall observed frequency 
	$\nu_{\mathrm{h}}$ of hydrogen-bond jump events per ion; and (3) the 
	fraction of these events that reverse themselves within a 50~fs 
	window.}
	\label{tab:hb_data}
\begin{tabular}
        {c|ccc|ccc|cc}
	\hline\hline
	& \multicolumn{3}{c}{\underline{\emph{Same SO$_4$}}}
	& \multicolumn{3}{c}{\underline{\emph{Different SO$_4$}}}
	& \multicolumn{2}{c}{\underline{\emph{Overall}}} \\
	~\emph{Temperature} (K)~ & 
	~\emph{\% tot.}~ & ~$\nu_{\mathrm{h}}$ (THz)~ & ~\emph{\% rev.}~ & 
	~\emph{\% tot.}~ & ~$\nu_{\mathrm{h}}$ (THz)~ & ~\emph{\% rev.}~ & 
	~$\nu_{\mathrm{h}}$ (THz)~ & ~\emph{\% rev.}~ \\
	\hline
	550 & 38 & 0.47 & 49 & 62 & 0.79 & 34 & 1.16 & 40 \\
	620 & 25 & 0.58 & 38 & 75 & 1.72 & 38 & 2.30 & 38 \\
	750 & 23 & 1.01 & 35 & 77 & 3.45 & 39 & 4.46 & 39 \\
	620 (fixed-SO$_4$) & 21 & 0.96 & 67 & 79 & 3.57 & 79 & 4.53 & 77 \\
	\hline\hline
\end{tabular}
\end{table*}

Unlike in the case of the chemical-bond dynamics, freezing the degrees of 
freedom of the SO$_4$ tetrahedra does not prevent reconfiguration of the 
hydrogen-bond network via bond breaking and forming. In fact,
Table~\ref{tab:hb_data} reveals that inhibiting SO$_4$ rotation actually 
enhances the frequency of hydrogen-bond breaking and forming, particularly for 
bonds exchanged between oxygens of neighboring tetrahedra. However, some degree
of rotation is required in order to visit a larger region of configuration
space and prevent repeated visitation of identical configurations, a vital 
stipulation for macroscopic proton transport. 

The point of separation of the hydrogen-bond autocorrelation curves for the 
fixed-SO$_{4}$ and fully mobile system, as seen in the inset of 
Fig.~\ref{fig:cbhb_autocorr}(b), can be interpreted physically as a
characteristic reversal time: if a newly formed hydrogen bond is to be 
accepted, enough SO$_{4}$ rotation must occur within the approximately 50~fs 
window to sufficiently alter and imbalance the energy landscape, thereby 
minimizing likelihood of back hopping. Accordingly, Table~\ref{tab:hb_data} 
shows that the observed fraction of hydrogen-bond network reorganization 
events that subsequently reverse themselves within this time period in the 
fixed-SO$_4$ simulation is more than double that of the fully mobile 
simulation at the same temperature (38\% versus 77\%).

\begin{figure}
	\centering
	\includegraphics[angle=270]{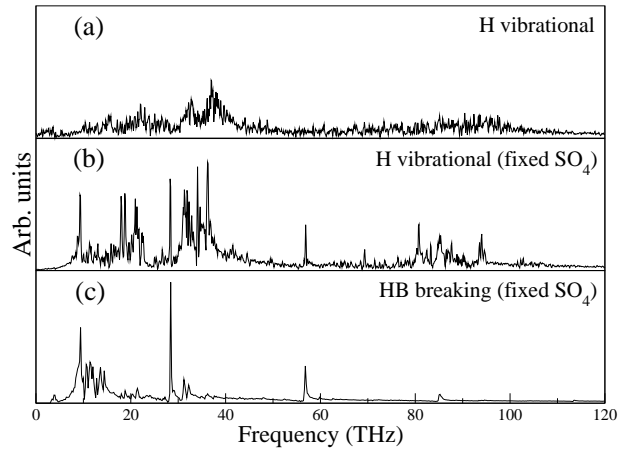}
	\caption{Vibrational density of states for the protons in (a) the 
	fully mobile and (b) the fixed-SO$_{4}$ simulations, along with
	(c) the hydrogen-bond switching frequency spectrum for the 
	fixed-SO$_{4}$ simulation. Data are from simulations at 620~K, and
	densities of states were obtained from a Fourier transform of the 
	appropriate velocity autocorrelation function.}
	\label{fig:frequencies_compare}
\end{figure}

A Fourier transform of the hydrogen-bond existence autocorrelation function 
$C_e(t)$ gives a good measure of the typical oscillation frequencies for the 
hydrogen-bond forming and breaking in the fixed-SO$_{4}$ simulation. 
Fig.~\ref{fig:frequencies_compare} compares this result to the vibrational 
density of states for the hydrogens in that simulation as well as in the fully 
mobile simulation. A comparison of Figs.~\ref{fig:frequencies_compare}(b) and 
(c) allows us to distinguish the hydrogen vibrational modes that are not 
directly connected to hydrogen-bond breaking from those that are. Those not 
linked to changes in the hydrogen-bond network are represented by clusters of 
broader peaks around 15--20~THz, 30--40~THz, and 80--95~THz. Of these, the two 
lowest-frequency clusters most likely represent bending modes, whereas the 
highest-frequency cluster contains stretching modes. The primary peaks 
associated with bond breaking and forming are a low-freqency peak near 9~THz 
and a fundamental second peak at 28~THz, along with its accompanying overtone 
peaks at higher frequencies. Since these peaks are completely suppressed in 
the result for the fully mobile simulation shown in 
Fig.~\ref{fig:frequencies_compare}(a), they represent the signature 
oscillations inhibited upon stabilization of new configurations by 
reorientations of the SO$_4$ tetrahedra. Notably, the half-period switching 
time represented by the low-frequency peak matches the characteristic reversal 
threshold obtained from Fig.~\ref{fig:cbhb_autocorr}. The higher-frequency 
peak at 28~THz is also evident in that same figure, appearing as shallow 
oscillations at short timescales. The existence of pronounced overtones for 
the 28~THz peak in both the vibrational and bond-breaking frequency spectra 
suggests significant anharmonicity in the potential for the H$\cdots$O bond.

Since the primary effect of suppressing SO$_{4}$ rotation is to encourage 
reversal of hydrogen-bond network reorientation phenomena rather than to 
inhibit such reorientations altogether, SO$_4$ rotation alone cannot 
satisfactorily account for the full network dynamics. Rather, we observe that 
the dominant mechanism for hydrogen-bond network reorganization involves 
hydrogen-bond transitions that are best described as rapid, discrete angular 
jumps between two stable states rather than as a smooth evolution driven by 
SO$_{4}$ tetrahedral rotations, as has generally been proposed previously. 
These two states correspond to different orientations for which the 
$\angle$S--O--H angle for chemically bonded hydrogens is maintained near the 
tetrahedral angle of 109.5$^\circ$. This phenomenon is depicted schematically 
in Fig.~\ref{fig:hopping}(c) and resembles the proposed diffusion mechanism 
in a recent simulation of liquid water\cite{laage06}.

\begin{figure}
	\centering
	\includegraphics[angle=270]{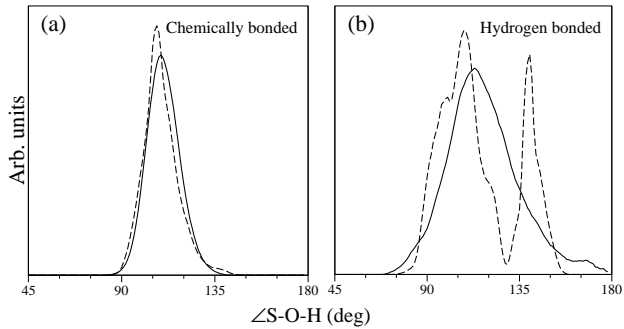}
	\caption{S--O--H angles for (a) chemically bonded and (b) 
	hydrogen-bonded protons in fully mobile (solid) and 
	fixed-SO$_{4}$ (dashed) simulations at 620~K.}
	\label{fig:soh_angles}
\end{figure}

Additional evidence for the angular hopping model appears in 
Fig.~\ref{fig:soh_angles}, which outlines distributions for certain 
geometrically relevant angles in both the fully mobile and fixed-SO$_4$
simulations at 620~K. The $\angle$S--O--H angles for chemically bonded protons 
have a relatively small spread and are peaked around the described tetrahedral 
geometry. Angles greater than 145$^\circ$ are not represented, indicating that 
the protons lie primarily on the surface of a cone centered on the S--O bond 
and with a half-angle of 65--70$^\circ$. Notably, the angular distribution 
does not change appreciably between the fully mobile and fixed-SO$_4$ 
simulations. Although hydrogen-bonded protons are generally less constrained, 
the $\angle$S--O$\cdots$H distributions for both simulations are still peaked 
near 110$^\circ$. However, for the fixed-SO$_4$ simulation, a second peak 
appears in the angular distribution at around 140$^\circ$ as a byproduct of 
the hydrogen-bond hops. The separation of the two peaks for the fixed-SO$_{4}$ 
case in Fig.~\ref{fig:soh_angles}(b) indicates that generally a net SO$_{4}$ 
tetrahedral reorientation of around 30$^\circ$, involving either the 
hydrogen-bond donor or acceptor tetrahedron, accompanies the hop to alleviate 
the lattice strain it induces. This value for the SO$_4$ angular rotation 
agrees well with what has been proposed in the literature
\cite{jirak87,merinov87,belushkin91a,belushkin91b}, but our resolution is 
insufficient to pinpoint which of the particular competing models is most 
likely to be correct. Such reorientation is also responsible for altering the 
potential energy surface to minimize back hopping. 

\begin{figure}
	\centering
	\includegraphics[width=3.375in]{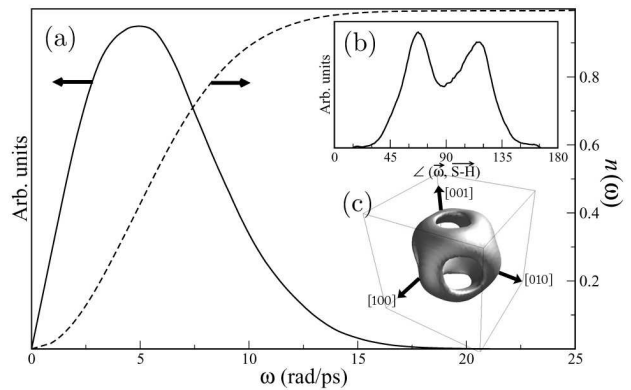}
	\caption{(a) Angular velocity profile for the SO$_{4}$ tetrahedra at 
	620~K (solid), along with the corresponding fraction of tetrahedra 
	$n(\omega)$ with velocities $\leq\omega$ (dashed). Inset (b) shows the 
	angular distribution of the sulfate tetrahedral axes of rotation 
	$\boldsymbol{\hat{\omega}}$ with respect to 
	$\mathbf{\hat{r}}_{\mathrm{SH}}$, and inset (c) shows the radial 
	distribution of the $\boldsymbol{\hat{\omega}}$ vectors in 
	space with respect to the primary crystallographic axes.}
	\label{fig:ang_vel_dist}
\end{figure}

\section{Rotation of SO$_{4}$ tetrahedra}

Since it has already been established that the 30$^\circ$ reorientation of the
SO$_4$ tetrahedra must take place within a 50~fs window to maximize the 
potential for non-reversing transitions, we can obtain 10.5~rad/ps as a 
back-of-the-envelope estimate for the minimum SO$_4$ angular velocity required 
to prevent reversal following a hydrogen-bond switch. The SO$_4$ angular 
velocities follow a Boltzmann distribution and are plotted in 
Fig.~\ref{fig:ang_vel_dist}. The plot reveals that velocities of this 
magnitude, although somewhat rare, are indeed accessible, representing about 
8--9\% of the SO$_4$ tetrahedra in the 620~K simulation at any given time. 

Fig.~\ref{fig:ang_vel_dist}(c) shows an isosurface of the angular velocity unit
vectors $\boldsymbol{\hat{\omega}}$ for the rotation of the SO$_4$ tetrahedra, 
averaged over all such groups in the 620~K simulation. Areas of high density 
therefore represent preferred axes of rotation, a clear structure for which is 
visible in the figure. These rotation axes do not align towards the chemically 
bonded hydrogen or its accompanying oxygen, as is evident from 
Fig.~\ref{fig:ang_vel_dist}(b). Instead, they orient along the edges of a cube 
rotated $\pi/4$ in the (001) crystal plane with respect to the conventional
unit cell, thereby correlating with the centers of the nearest-neighbor 
tetrahedra. The SO$_4$ rotational orientations thus appear to be governed by 
the locations of nearby SO$_4$ tetrahedra rather than the location of the
locally bonded hydrogen or the orientation of its corresponding chemical bond.
In other terms, rotational axes align along $\mathbf{\hat{r}}_{\mathrm{SS}}$ 
rather than $\mathbf{\hat{r}}_{\mathrm{SH}}$ or 
$\mathbf{\hat{r}}_{\mathrm{SO}}$.

\begin{figure}
	\centering
	\includegraphics[angle=270]{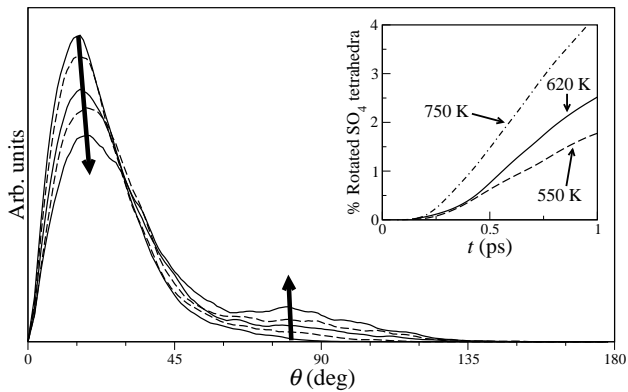}
	\caption{Distribution of angular distances traveled by sulfate 
	tetrahedra as a function of time at 620~K. Successive curves represent 
	values at 0.5, 1, 2, 3, and 5~ps. The inset gives a measure of the 
	fraction of tetrahedra with greater than 70$^\circ$ rotation from 
	their initial positions as a function of time at 550~K, 620~K, and 
	750~K.}
	\label{fig:so4_rotation_time}
\end{figure}

In view of the fact that angular jumps alone, stabilized by small, rapid 
tetrahedral reorientation events of about $30^\circ$, account for much of the 
hydrogen bond network dynamics, it is desirable to analyze and requantify the 
relative contribution of larger-scale rotational motion. 
Fig.~\ref{fig:so4_rotation_time}, which shows the averaged distributions of  
angular distances traveled by SO$_4$ tetrahedra as a function of time, 
illustrates one effect of slower rotational dynamics. There is evidence of the
appearance of a second peak in the distributions representing a new, stable 
equilibrium configuration at 75--80$^\circ$ rotation with respect to the 
original tetrahedral orientation. This peak begins to manifest in statistically
measurable quantities only after 250~fs (see inset of figure), making the 
quickest of such reorientation events several times slower than the timescale 
of the fast 30$^\circ$ reorientation event described previously. We note that 
the faster dynamics with the smaller reorientation angle cannot be 
distinguished in Fig.~\ref{fig:so4_rotation_time} since it is buried within 
the first peak, which primarily depicts the fast librational modes. 

\begin{figure}
	\centering
	\includegraphics[angle=270]{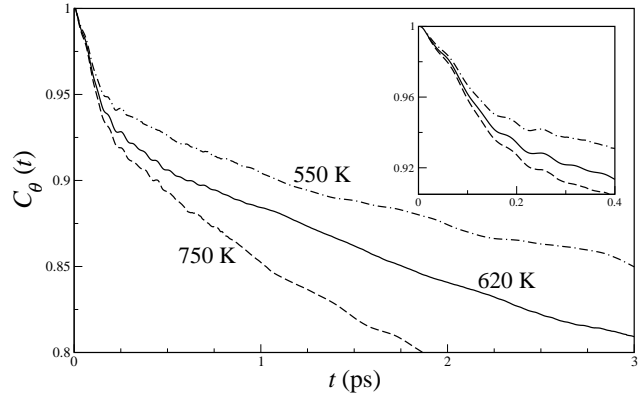}
	\caption{Autocorrelation function for angular distance traveled 
	between SO$_{4}$ tetrahedral orientations separated by time $t$.}
	\label{fig:angdist_autocorr}
\end{figure}

Additional confirmation of the existence of multiple distinct timescales for 
the tetrahedral orientation can be seen in Fig.~\ref{fig:angdist_autocorr}, 
which portrays the angular time autocorrelation of tetrahedral configurations. 
We calculate this quantity according to $C_{\theta}(t)=\langle
\mathbf{\hat{r}}_{\mathrm{SO}}(t) \cdot \mathbf{\hat{r}}_{\mathrm{SO}}(0) 
\rangle$, producing a measure of the cosine of the average angular distance 
traveled by an S--O unit vector $\mathbf{\hat{r}}_{\mathrm{SO}}$ as a function 
of time. As before, we maximize our available statistics by averaging 
$C_{\theta}(t)$ over all available time intervals of length $t$ in the 
simulation. The slower rotation events manifest themselves as a quasi-linear 
decay in the autocorrelation at longer times. At short times ($<250$~fs), the 
slope is appreciably steeper, indicating faster dynamics on average. 
Expectedly, the separation between these two regimes agrees with the timescale 
of the emergence of the slow rotation in Fig.~\ref{fig:so4_rotation_time}. In 
addition, a shoulder indicating the timescale of a rotation to a nearby local 
minima is clearly distinguishable at around 50--60~fs, in agreement with our 
previous indicators of fast dynamics on that scale. This shoulder repeats 
itself as periodic oscillations and is also detectable as a peak in the 
Fourier transform (not shown) of the curve in Fig.~\ref{fig:angdist_autocorr}. 
The oscillations also span the intermediate region in which rotations of both 
short and long timescales are manifest before they are lost in the slower 
dynamics at longer times. We further note that changes in temperature have no
appreciable effect on the timescale of the fast rotation, as measured by the 
locations of the shoulders and oscillations in the figure. This is consistent 
with a picture in which the reorientation is connected to a hydrogen-bond hop 
and therefore has an almost negligible rotational barrier.

The anisotropy of the diffusion tensor for CsHSO$_{4}$ has been documented
experimentally\cite{sinitsyn91} and is a geometric consequence of the greater 
angular distance that must be traversed by a diffusing proton traveling across 
two sulfate layers along the $[$001$]$ or $[$00$\overline{1}]$ directions 
(117$^\circ$) compared to a similar journey in a $\{$001$\}$ plane 
(78$^\circ$). Our results confirm that diffusion parallel to the (001) plane
dominates: the directional mean-square displacement of the hydrogen atoms 
rises 2.5 to 5 times faster along the $[$010$]$ or $[$100$]$ directions than 
along the $[$001$]$ direction, with higher temperatures favoring higher 
anisotropy. This disparity suggests a different dominant mechanism for 
diffusion along a $\langle$001$\rangle$ direction, in part because the angular 
distance is too great to be easily accommodated by the described hydrogen bond 
hopping mechanism, and in part because the corresponding angular velocities 
that would have to be reached by the sulfate tetrahedra to prevent backhopping 
in such a scheme are unreasonably high. Instead, we find that slower SO$_{4}$ 
rotation plays the dominant role in overcoming the larger barrier, in line 
with the more traditional model of proton transport in CsHSO$_{4}$ (depicted in
Fig.~\ref{fig:hopping}(b)). In this case, the dynamics are slow enough that 
the timescale of the large-scale rotation is no longer a hindrance. We further 
note that the approximately 40$^{\circ}$ difference in the angular distance 
that must be traveled in the $[$001$]$ direction with respect to a similar 
journey in the (001) plane, combined with the aforementioned 
30$^{\circ}$ SO$_{4}$ reorientation, satisfactorily accounts for the 
appearance of the second peak in Fig.~\ref{fig:so4_rotation_time} to within a 
rough estimate. This suggests that for diffusion in the $[$001$]$ direction, 
the slow rotation is probably followed by a rapid hydrogen-bond hop, although 
the former clearly determines the timescale.

\section{Hydrogen-bond network topology}

We have also analyzed the basic topology of the hydrogen-bond network.
Table~\ref{tab:so4_prob} shows the relative probabilities of various SO$_{4}$ 
bonding configurations at 550~K, 620~K, and 750~K, organized according to the 
number of hydrogen bonds donated ($N_d$) and accepted ($N_a$) by the SO$_{4}$ 
tetrahedron. In our definition, a donated bond is formed between a local 
chemically bonded hydrogen and an oxygen on a neighboring tetrahedron; an 
accepted bond is a hydrogen bond formed between a local oxygen and a hydrogen 
that is chemically bonded to a neighboring tetrahedron. For an ideal 
one-dimensional network, one would expect all tetrahedra to have 
$N_d$=$N_a$=$1$. In our simulations, we observe this type of ordinary link only 
41--61\% of the time, suggesting the actual network topology is much more 
complicated. Other highly probable configurations include one with 
$(N_a, N_d)=(1, 0)$, which can be thought of as a terminator in the hydrogen 
bond network (18--26\%); and one with $(N_a, N_d)=(1, 2)$, which can be 
interpreted as a network branching point (8--14\%). According to the table, 
there is a great deal of variability in the number of hydrogen-bond donors 
$N_d$, but configurations with $N_a$$\not=$$1$ are comparatively rare. The 
primary effect of increasing the temperature seems to be a decrease in the
number of ordinary linear network links with $N_d$=$N_a$=$1$ in favor of 
network terminators with $(N_a, N_d)=(1, 0)$, resulting in a more nodal 
network. Some of the rare (but nonetheless statistically significant) 
configurations are signatures of Grotthus-type jumps in progress: immediately 
following a standard chemical-bond jump from one SO$_4$ tetrahedron to 
another, nucleated at a link in an ordinary linear chain, the source 
tetrahedron registers a topological configuration of the form 
$(N_a, N_d)=(2, 0)$, whereas the H$_2$SO$_4$ destination tetrahedron
acquires a configuration of the form $(N_a, N_d)=(0, 2)$. 

\begin{table*}
	\caption{Observed relative probabilities (\%) of SO$_4$ bonding 
	configurations, organized according to the number of hydrogen bonds 
	accepted (rows) and the number donated (columns) by the SO$_4$ 
	tetrahedron. Data are from simulations at 550/620/750~K.}
	\label{tab:so4_prob}
\begin{tabular*}{0.75\textwidth}
        {@{\extracolsep{\fill}}c|cccc}
	\hline\hline
	~\emph{\% Prob.}~ & $N_d=0$ & $N_d=1$ & $N_d=2$ & $N_d=3$ \\
	\hline
	$N_a=0$ & 2~/~3~/~7 & 3~/~4~/~6 & 1~/~1~/~2 &
1~/~1~/~1 \\
	$N_a=1$ & 18~/~22~/~26 & 61~/~53~/~41 & 11~/~15~/~14 &
1~/~1~/~1 \\
	$N_a=2$ & 1~/~1~/~2 & $<$1~/~$<$1~/~1 & 0~/~0~/~0
& 0~/~0~/~0 \\
	\hline\hline
\end{tabular*}
\end{table*}

\begin{table*}
	\caption{Average lifetimes (fs) of SO$_{4}$ bonding configurations,
	organized according to the number of hydrogen bonds accepted
	(rows) and the number donated (columns) by the tetrahedron. Data are
	from simulations at 550/620/750~K.}
	\label{tab:so4_lifetimes}
\begin{tabular*}{0.75\textwidth}
        {@{\extracolsep{\fill}}c|cccc}
	\hline\hline
	\emph{Lifetime (fs)} & $N_d=0$ & $N_d=1$ & $N_d=2$ & $N_d=3$ \\
	\hline
	$N_a=0$ & 48~/~53~/~55 & 47~/~43~/~42 & 21~/~19~/~25 & 34~/~21~/~24 \\
	$N_a=1$ & 176~/~159~/~117 & 238~/~169~/~105 & 128~/~119~/~75 &
81~/~92~/~54 \\
	$N_a=2$ & 27~/~20~/~24 & 20~/~17~/~20 & 0~/~0~/~0 & 0~/~0~/~0 \\
	\hline\hline
\end{tabular*}
\end{table*}

\begin{figure}
	\centering
	\includegraphics[width=3.375in]{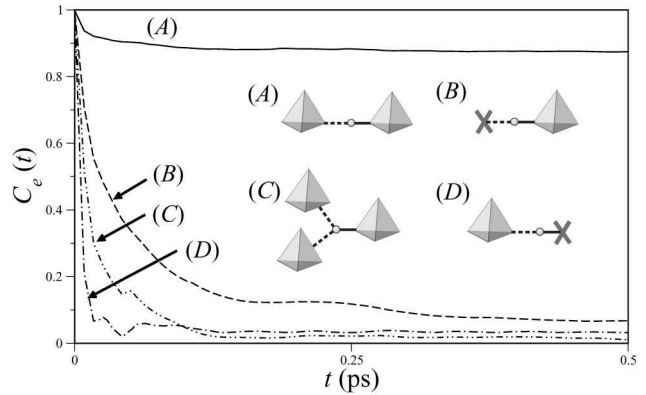}
	\caption{Existence autocorrelation functions for four of the most
	likely tetrahedral bonding configurations. In the schematic diagrams,
	a solid line represents a chemical bond, a dashed line represents a 
	hydrogen bond, and an ``X'' indicates the absence of a bond.}
	\label{fig:defect_autocorr}
\end{figure}

Table~\ref{tab:so4_lifetimes} lists the average lifetimes of the local
topologies listed in Table~\ref{tab:so4_prob}. Although these values are
averages and do not account for the complete distribution of possible 
lifetimes as do the autocorrelation curves of Fig.~\ref{fig:defect_autocorr}, 
they are nonetheless useful for purposes of qualitative comparison. Lifetimes
are generally well correlated with relative frequencies, with hydrogen bonding 
to a single secondary tetrahedron ($N_a$=1) acting as a stabilizing force. As 
the temperature increases, the lifetimes of configurations with 
$N_a$$\not=$$1$ are affected very little, but we observe a sharp systematic 
decline in nearly all configurations with $N_a$=$1$. Notably, this trend does 
not always follow that of the relative frequencies in Table~\ref{tab:so4_prob}. 
For example, network terminators with $(N_a, N_d)=(1, 0)$ exhibit a decrease 
in average lifetime but an increase in their overall commonality, suggesting a 
more nodal but also more dynamic network at high temperatures.

Tables~\ref{tab:so4_prob} and \ref{tab:so4_lifetimes} neglect hydrogen bonds 
to the same SO$_4$ tetrahedron as the host, in accordance with our original 
definition of the hydrogen bond. However, if we relax the hydrogen-bonding 
restriction requiring bonds to be between oxygens of different tetrahedra, we 
find that these ``self-hydrogen bonded defects'', although short-lived, are 
nonetheless relatively common, representing about 3\% of the total O$\cdots$H 
interactions at 620~K. Moreover, we find that hydrogens involved in chemical 
and hydrogen bonds to the same SO$_{4}$ tetrahedron do not permit simultaneous 
hydrogen bonding to oxygens of neighboring tetrahedra, meaning these complexes 
function as terminators for the hydrogen-bond network chains. Also, the 
average velocity of oxygens in self-hydrogen bonded SO$_4$ complexes is 
consistently about 10\% higher on average than in ordinary complexes with 
$N_d$$=$$N_a$$=$$1$. This suggests that underbonding lessens the degree of 
constraint and enhances oxygen mobility in such units, likely aiding further 
reorganization of the hydrogen-bond network.

In addition to examining the network topology in terms of connectivity between
neighboring SO$_{4}$ tetrahedra, one can obtain a slightly different
topological gauge by looking at the number of chemical and hydrogen bonds 
formed by a single proton. Fig.~\ref{fig:defect_autocorr} shows the existence 
autocorrelation curves for four of the most common topologically distinct
bonding configurations for a proton. These curves give an idea of the 
characteristic decay times for each configuration. The result for a standard
bonding configuration, in which a proton forms one chemical and one hydrogen
bond (\emph{A}) is shown for timescale reference. We note that topologies with 
protons forming a single chemical bond but no hydrogen bond (\emph{B}) are
surprisingly stable, with a characteristic decay time that is relatively large
on the timescale of both the hydrogen-bond hopping and its corresponding 
strain-relaxing SO$_{4}$ reorientation. Defects of this class are less 
constrained and more mobile than their ordinary counterparts, and as in the 
case of the self-hydrogen bonded complexes, the increased mobility facilitates
network reorganization much more readily. Network-branching configurations 
with multiple hydrogen bonds (\emph{C}) play a more direct role in the 
reorganization of the hydrogen-bond network but have only intermediate
stability. Configurations with no chemical bond (\emph{D}) are extremely 
short-lived.

Techniques involving the graph-theoretic adjacency matrix (see Appendix) also 
offer a convenient way of characterizing the topology of the overall network 
and extracting configurations most likely to induce a diffusive event. In 
particular, we are able to further classify the network topology in terms of 
\emph{rings}, meaning some part of the network ultimately connects back to 
itself in a closed loop; and \emph{chains}, meaning the network either remains 
linear or branches, with the restriction that any two network vertices are 
connected by exactly one unique directed path (i.e., a graph-theoretic tree). 
The specifics of our classification algorithm are described in the Appendix. 
It should be noted that such a dichotomy requires classification of every node 
as either a chain or a ring but does not allow any given node to be doubly 
counted as belonging to both categories. Table~\ref{tab:so4_rings_chains} 
lists the likelihood of finding a tetrahedron in various ring and chain 
topologies in an ordinary simulation timestep versus a timestep immediately 
preceding a chemical- or hydrogen-bond jump event. 

For ordinary configurations not involving a jump event, the network favors
rings over chains at 550~K, whereas the trend is reversed at 750~K. The 
intermediate temperature of 620~K is a topological transition zone and shows a 
marked increase in configurations simultaneously containing both rings and 
chains. There is also a notable decrease in the average length of a chain and a
slight decrease in the average size of a ring at 620~K compared to the other 
temperatures. This is a further indication that the network is midway in a
transition process from primarily rings to primarily chains, as a network 
configuration containing both would tend to inhibit the growth of either one 
at the expense of the other.  

Our findings indicate that at all temperatures, the presence of topological 
chains has a dramatic effect in enhancing the likelihood of occurence of 
either sort of jump event. Conversely, the presence of rings strongly inhibits 
jumping. The trend is much more pronounced when one examines frames containing 
either rings or chains as the only topological species: in frames preceding a 
chemical-bond jump, we see a 23--24\% increase in likelihood of the frame to
contain exclusively chains and a corresponding 13--23\% decrease in its 
likelihood to contain exclusively rings, as compared to a frame in an ordinary
non-jumping configuration. This difference is especially pronounced at higher
temperatures. Nonuniform configurations containing both rings and chains also 
show an overall decrease in jump likelihood. The trends in the chain and ring 
data also evidence that from a purely topological perspective, a configuration 
favorable for a hydrogen-bond jump lies midway between an ordinary 
configuration and one favorable for a chemical-bond jump. We conclude that the 
network ring and chain topology is a good indicator of both hydrogen- and 
chemical-bond jump likelihood and is substantially more effective in that 
capacity than the measure of the oxygen-oxygen distance across the 
O--H$\cdots$O complex.

It is worthwhile mentioning that periodic boundary conditions and limited 
supercell sizes have two major topological consequences that must be 
considered in any analysis: first, they decrease the maximum length of chains 
that can be formed; and second, they tend to artificially inflate the number 
of smaller rings, since creation of periodic images tends to wrap the network 
back onto itself prematurely. As such, the values in 
Table~\ref{tab:so4_rings_chains} should not be taken as absolutes, but 
qualitative comparisons are nonetheless useful and relevant.

\begin{table*}
	\caption{Observed relative probabilities (\%) for various hydrogen-bond 
	network topologies in an ordinary timestep, compared with similar
	quantities for timesteps immediately preceding a chemical- or 
	hydrogen-bond jump event. Also listed are the relevant average ring 
	and chain sizes for frames where those topologies exist. Ring sizes 
	are calculated based on the number of tetrahedra involved in the ring, 
	whereas chain sizes denote the maximum individual branch length within 
	the graph-theoretic tree structure. Data are from simulations at 
	550/620/750~K.}
	\label{tab:so4_rings_chains}
\begin{tabular}
        {l|c|c|c}
	\hline\hline
	~\emph{Description}~ & ~~~~~~\emph{No jump}~~~~~~ & ~\emph{Hydrogen-bond jump}~ &
~\emph{Chemical-bond jump}~\\
	\hline
	~Contains rings~ & 71~/~78~/~52 & 64~/~72~/~49 & 47~/~55~/~28 \\
	~Contains chains~ & 60~/~74~/~70 & 69~/~81~/~75 & 83~/~87~/~84 \\
	~Contains only rings~ & 40~/~26~/~30 & 31~/~19~/~25 & 17~/~13~/~16 \\
	~Contains only chains~ & 29~/~22~/~48 & 36~/~28~/~51 & 53~/~45~/~72 \\
	~Contains both rings and chains~ & 31~/~52~/~22 & 34~/~52~/~24 &
29~/~42~/~12 \\
	\hline
	~Average ring size~ & 5.6~/~4.5~/~4.7 & 5.3~/~4.5~/~4.6 & 4.9~/~4.5~/~4.5 \\
	~Average chain size~ & 6.9~/~5.4~/~6.2 & 6.9~/~5.4~/~6.0 & 6.8~/~5.6~/~6.0 \\
	\hline\hline
\end{tabular}
\end{table*}

\section{Proton kinetics and the isotope effect}

We can estimate the general three-dimensional proton self-diffusion coefficient 
from the mean-square displacement (MSD) using the Einstein relation:
$D_{\mathrm{H}}=\lim_{t\rightarrow\infty}\frac{1}{6t}\left\langle
\mathrm{MSD}(t)\right\rangle$. Using this method, we estimate $D_\mathrm{H}$ 
in our simulations to be 1.8--3.5$\times10^{-6}$ cm$^2/$s over the 550--750~K 
temperature range, about an order of magnitude greater than extrapolations of 
experimental measurements to the same range (1.9--3.4$\times10^{-7}$ cm$^2/$s)
\cite{belushkin92}. However, despite the error in the magnitude of the 
calculated diffusion coefficients, we nonetheless observe the proper scaling 
of $D_\mathrm{H}$ with temperature, indicating correct calculation of the 
energetic barriers for diffusion.

Multiple factors may contribute to the discrepancy of the calculated diffusion 
coefficients with experiment. First, calculations of these coefficients are 
notoriously difficult to converge, particularly when diffusivities are small 
and statistics are limited. One possible gauge of accuracy in the calculation 
can be achieved by comparing the result to the same quantity found using the 
Green-Kubo relation. The Green-Kubo method is based on the velocity 
autocorrelation function rather than the mean-square displacement; however, 
this method generally exhibits even poorer convergence in the absence of 
excellent statistics. Instead, we estimate the accuracy of our diffusion
coefficients based on twice the standard error in the determination of the 
mean-square displacement slope, which results in an error estimate of 
$\pm$15\%.

Second, we have already mentioned the difficulty in accounting for correlation
and reversal of collective sequences of jump events. In the limit of the short 
length- and timescales accessible by first-principles methods, it is expected
that any diffusive correlations that persist over large regions of either time
or space would be lost, resulting in inflated diffusion statistics. Similarly, 
jump events that remain localized due to high incidences of back-hopping and 
correlation are counted as contributors to diffusion on the timescale 
accessible to our simulations, whereas these would not appear in a macroscopic 
measurement.

We have also mentioned the artificially high number of small rings due to 
periodic boundary conditions as a potential consequence of small supercell
size. Table~\ref{tab:so4_rings_chains} suggests a possible correlation 
between smaller average ring size and a higher propensity to jump, particularly
at the lower temperature, meaning our finite size effects could result in a 
measurable increase in jump statistics and diffusion coefficients. However, 
any such decrease in average ring size from the periodic boundary conditions 
is likely to be accompanied by an increase in the overall number of rings. 
Since we have established the overall propensity for ring existence to inhibit 
hydrogen- and chemical-bond dynamics, the competition between these two 
effects should attenuate any potential impact on the macroscopic properties.

It should be noted that we are neglecting any quantum behavior of the protons. 
However, an analysis of experiments on CsDSO$_{4}$ suggests the isotope effect 
is relatively small\cite{kreuer96,sinitsyn91}. In addition, theoretical work 
on the topologically similar material KDP\cite{tosatti02} concluded that the 
predominant effect of quantum delocalization of the protons was limited to 
structural considerations, in that it decreased the H$\cdots$O--H distance and 
consequently also the lattice parameter. The KDP analysis is also consistent 
with experimental comparisons of CsDSO$_{4}$ and CsHSO$_{4}$
\cite{merinov87,belushkin91a}. 

A closer examination of the specific rate-limiting mechanisms covered in our
analysis of proton diffusion in CsHSO$_4$ provides additional insight into the
lack of any substantial isotope effect. Our findings indicate that the 
chemical-bond dynamics are much faster than previous analyses have suggested, 
of the same order as the hydrogen-bond dynamics. Rather, the primary 
limitation is manifest in the dynamics of the SO$_4$ tetrahedra, since these 
appear to govern the reversal rates of both chemical-bond jumping and 
hydrogen-bond hopping. Accounting for proton tunneling across the 
O--H$\cdots$O double-well potential would therefore have little effect on the
overall jump statistics, since the mobility of the SO$_4$ groups is classically
controlled. In addition, we have established the importance of the chain and 
ring topology of the hydrogen-bond network in promoting diffusive events.
However, changes in the topology are promoted by two factors---hydrogen-bond 
hopping and slow rotation of the SO$_4$ tetrahedra. Slow SO$_4$ rotation is 
clearly a classical phenomenon, and hydrogen-bond hopping is coupled to a 
30$^{\circ}$ reorientation of the heavy SO$_4$ tetrahedron, meaning its 
dynamics are also ultimately classical.

\section{Conclusions}

We have presented a detailed analysis of proton dynamics in superprotonic 
Phase-I CsHSO$_4$ based on first-principles molecular dynamics simulations. 
Our results confirm that the chemical-bond dynamics are dominated by local 
Grotthus-style hops which propagate successively along the hydrogen-bond 
network backbone. Individually, these hops are comparatively frequent, 
pointing to a low diffusion barrier, but the net effective rate of the 
chemical-bond dynamics is limited by an anomalously high rate of jump 
reversal. We find that the propensity for such forward- and back-hopping along 
the O--H$\cdots$O complex is in turn heavily influenced by the dynamics of the 
SO$_4$ tetrahedra rather than by static local geometry alone.

We have also shown that the dynamics of the hydrogen-bond network are 
dominated by fast, discrete angular jumps between neighboring oxygens rather
than by slow rotations of the SO$_4$ tetrahedra. Such jumps occur with greater 
frequency between oxygens belonging to different SO$_4$ tetrahedra than 
between oxygens of the same tetrahedron, by a factor of two or three. The
hydrogen-bond jumps are accompanied by an approximately 30$^{\circ}$ 
reorientation of the participating SO$_4$ tetrahedra to alleviate the lattice
strain induced by the hop, thereby minimizing the likelihood of jump reversal.
We have isolated a window of 50~fs for successful completion of this ``fast'' 
reorientation event and showed that it exists independently of a second, 
slower reorientation mechanism, operating on a timescale at least five times 
greater than its counterpart. The slower mechanism amounts to ordinary SO$_4$ 
rotation on a longer timescale, and we propose this to be the dominant 
hydrogen-bond network reorientation mechanism for diffusion along the 
$[$001$]$ direction, for which angular hops are significantly more difficult 
and less frequent, owing to the anisotropy of the CsHSO$_4$ lattice.

Our topological analysis of the hydrogen-bond network revealed a significant
number of branching and network-terminating nodes, indicating a substantial
deviation from linearity, particularly at higher temperatures. We postulate
that the underbound network terminators play a role in network reconfiguration
by aiding SO$_4$ rotational mobility. Graph-theoretic methodology offered a way
to isolate chains and rings as dominant topological features in the network,
and we discovered that the presence of chains and the absence of rings is a
substantial predictor of likelihood for either a hydrogen- or chemical-bond
jump event to occur. We propose that our topological analysis could be
easily extended to similar well-defined, hydrogen-bonded network solids.

Finally, we apply our analysis to offer an explanation for the lack of a 
significant isotope effect in the CsHSO$_4$/CsDSO$_4$ system. In particular,
we tie both the chemical- and hydrogen-bond dynamics to the classical dynamics 
of the SO$_4$ tetrahedra and argue that the inclusion of proton quantum 
tunneling should play a relatively minor role in the rate-limiting steps of 
the diffusion mechanism.

\section{Acknowledgments}

Funding for this work has been provided by the U.S. Department of Energy CSGF 
and MURI Grant DAAD 19-03-1-0169. Calculations have been done using the 
Quantum-ESPRESSO package\cite{espresso} on computational facilities provided 
through NSF grant DMR-0414849. 

\section{Appendix: Graph-theoretic methodology}

Here we offer a detailed account of the graph-theoretic methods used to
calculate the hydrogen-bond network topology. We represent the network as a
directed graph with the edge vector pointing along the O--H$\cdots$O bond
direction; that is, from the sulfate tetrahedron acting as the hydrogen-bond 
donor in the complex to the sulfate tetrahedron acting as the hydrogen-bond 
acceptor. The adjacency matrix $A_{ij}$ can then be constructed as an 
$N\times N$ matrix, where $N$ is the number of tetrahedra in the unit cell and
the indices $i$ and $j$ run over the donor and acceptor sulfate groups, 
respectively:
\[ A_{ij}=
\left\{ \begin{array}{l}
1 \mathrm{~if~there~exists~a~direct~link~} i\rightarrow j\\
0 \mathrm{~otherwise} \end{array} \right.\]
The diagonal elements $A_{ii}$ are set to zero. Topological characterization 
of a single node is then a straightforward process of performing a row sum to 
get the number of nodes to which it donates ($N_d$) and a column sum to get 
the number of nodes donating to it ($N_a$). It is also easy to categorize jump 
events by analyzing the difference of the adjacency matrices of successive 
timesteps.

To determine ring connectivity and size, we exploit the property of adjacency
matrices\cite{godsil01} that element $(i,j)$ of $A^{n}$ gives the number of 
unique directed pathways from $i$ to $j$ of length $n$. We take the size of 
the ring containing the $i^{th}$ node to be the lowest value of $n$ in the 
interval [2, $(N$-$1)$] for which the diagonal element $A_{ii}^{n}\neq 0$. If 
$A_{ii}^{n}=0$ for all $n$ in the interval, the node is not considered part of 
a ring.

Deriving chain sizes is somewhat more complex, since we must account for
multiple branching topologies and for topological mixtures of rings and 
chains. We first decompose the network into clusters of unconnected subgraphs. 
This is done using the connectivity matrix $C_{ij}$, which has the property 
that $C_{ij}$=$1$ if there is a path \emph{of any length} connecting nodes 
$i$ and $j$. We form the symmetric $C_{ij}$ from $A_{ij}$ using Warshall's 
algorithm\cite{warshall62}.

We proceed to construct a matrix $S_{ij}$ that contains the shortest path 
between each pair of nodes $(i,j)$, $i\neq j$, by taking the lowest value of 
$n$ in the interval [2, $(N$-$1)$] for which $A_{ij}^{n}\neq 0$. We can then 
take the maximum value of $S_{ij}$ across the columns to get a row array of 
maximum path lengths for chains originating at the $i^{th}$ node. The chain 
size is then determined by finding the maximum value of the resultant array 
over the nodes contained in each connected cluster, which can be easily 
determined by parsing $C_{ij}$. Finally, we add the restriction that none of
the links in the chain can themselves be members of rings. This prevents 
counting of chains that are fictitiously long due to intermediate or 
terminating rings and ensures a clear separation between ring and chain
topologies. The resulting chain size is then calculated as the maximum span of 
the graph-theoretic tree.

Very large networks may necessitate more efficient algorithms due to the
expense of calculating $A^{N-1}$. However, the system sizes in our study are 
sufficiently small to readily allow calculations using the described method.

\end{document}